\documentclass{optica-article}
\journal{opticajournal}

\articletype{Research Article}
\usepackage[utf8]{inputenc}
\usepackage{graphicx, animate}
\usepackage{lineno}
\usepackage{ulem}
\usepackage{appendix}
\usepackage{colortbl}
\usepackage{xcolor}
\usepackage{tabularray}
\usepackage{float}
\renewcommand{\thepage}{\arabic{page}}
\begin{document}

\title{Versatile Metamaterial: Exploring Symmetry-Protected Mode Resonances for Multi-Task Functionality}

\author{Souhaila Boublouh,\authormark{1} Miguel Suarez,\authormark{1} Gao Feng,\authormark{2,3} Abderrahmane Belkhir,\authormark{4}, Abdelkrim Khelif,\authormark{3,5} and Fadi I. Baida\authormark{1,*}}
\address{\authormark{1}Institut FEMTO-ST, UMR 6174 CNRS, D\'epartement d'Optique P. M. Duffieux, Universit\'{e} de Franche--Comt\'{e}, 25030 Besan\c{c}on Cedex, France\\
\authormark{2}University of Zhejiang, Hangzhou, Zhejiang, China\\
\authormark{3}Institut FEMTO-ST, UMR 6174 CNRS, D\'epartement MN2S, Universit\'{e} de Franche--Comt\'{e}, 25030 Besan\c{c}on Cedex, France\\
\authormark{4}Laboratoire de Physique et Chimie Quantique, Universit\'{e} Mouloud Mammeri, Tizi-Ouzou, Algeria\\
\authormark{5}College of Science and Engineering, Hamad Bin Khalifa University, Doha, Qatar}
\email{\authormark{*}fbaida@univ-fcomte.fr}

\begin{abstract}
In this article, we present an experimental study supported by numerical modeling showing the possibility of exciting Symmetry-Protected Bound states In the Continuum (SP-BICs) in a 1D silicon grating fabricated on a lithium niobate substrate in both transverse electric and transverse magnetic polarization states of the incident illumination. This leads to different resonances in the transmission spectra with large quality factors up to $10^6$ and a significant electric/magnetic field enhancement up to $10^5$ opening the way to the exploitation of this structure for different sensing applications (biological,  electromagnetic, thermal...) but also to nonlinear applications such as the generation of second harmonic, in addition to electro- and acousto-optic modulation.
\end{abstract}

\section{Introduction}
The enhancement of the light-matter interaction is a key point to exacerbate the optical response of a given structure leading to an increase in its performance \cite{lightmatterinteractions,lightmatter2}. To this end, two conditions must be fulfilled: first, the increase in the lifetime of photons within the structure corresponding to the excitation of resonances with a high-quality factor (Q-factor), and second, the resonance must provide a strong confinement of the electromagnetic field at the sub-wavelength scale (small mode volume). The combination of these two conditions leads to resonance with a large Purcell factor \cite{purcellfactor}. Otherwise, plasmonic resonances \cite{plasmonics} are probably not suitable in this context due to the involved intrinsic absorption of used metals. Thus, nowadays, most research teams focus on dielectric structures \cite{All-dielectricmeta} fulfilling the two conditions mentioned above. This can be achieved quite easily in some specific cases, such as the excitation of a frozen mode within a photonic crystal \cite{frozenmode}, a Fano-like resonance mode \cite{FanoResonanceQualityFactor}, a Guided Mode Resonance (GMR) \cite{GMR} or, more recently, a Bound state In the Continuum (BIC) \cite{BICs}. In the latter case, these modes can be interpreted as non-radiative, resulting from destructive interference between two or more leaky modes, and are therefore not visible in optical spectra \cite{ReviewBICsNature}. A typical example of such modes is the family of Symmetry Protected Modes (SPMs) \cite{hoblos2020excitation}. The non-radiative character (dark modes) corresponds to resonance with an infinite Q-factor \cite{BICs}. Symmetry breaking allows to excite such modes with a Q-factor all the more important as the degree of symmetry breaking is low \cite{SPBICExperimental}. The symmetry is often geometrical (with respect to a specific axis or plane combined with an illumination having the same symmetry) but can also be of electromagnetic origin (the geometrical symmetry is preserved but with illumination having a different axis/plane of symmetry than the grating) \cite{ReviewBICsNature}.      
 BICs take part in many applications due to the extreme light confinement on a small scale, and their high sensitivity toward small external variations such as the refractive index of the medium. Typical applications include the enhancement of optical nonlinearities, where BICs can boost the efficiency of various nonlinear optical phenomena, such as THG in Si metasurfaces, or SHG in non-centro-symmetric materials \cite{koshelev2019nonlinear,liu2019high}, 
 
In this paper, we propose a 1D subwavelength grating with a very simple geometry that supports multiple BICs and SPMs in the near-infrared range. The structure has been fabricated and characterized under different illumination conditions showing mutiple resonances with a nearly high Q-factor. Numerical simulations by FDTD (Finite Difference Time Domain) are in great agreement with the experimental data of transmission spectra. The calculations also show that these high Q-factor resonances are accompanied by a significant enhancement of the electric or/and magnetic fields, which demonstrate strong electromagnetic field confinement either in the superstrate (air), in the substrate (lithium niobate), or in the structured medium itself. This suggests diverse applications such as biological detection (index variations), in electro-, acoustic- or pyroelectric modulation, with the same component operating at different wavelengths and for different illuminations.

\section{Proposed geometry}
\begin{figure}[H]\centering
{\includegraphics[clip,width=0.9\textwidth]{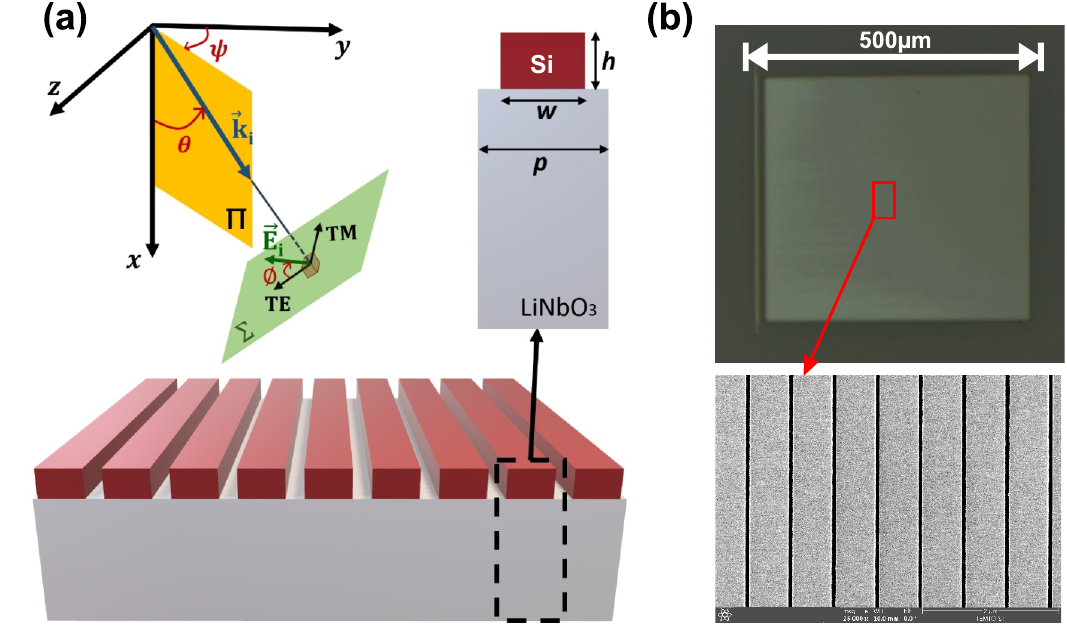}}
\caption{(a) Illustration of the proposed grating design featuring Si-pillars atop an X-cut LN substrate. The grating parameters include the period $p$, pillar gap $w$, and pillar height $h$. Illumination characteristics are determined by three angles: the angles of incidence $\theta$, azimuthal angle $\psi$, and the polarization angle $\phi$. $\Pi$ represents the plane of incidence, while $\Sigma$ denotes the wavefront. (b) Scanning electron microscope (SEM) image captured from the top view of the fabricated grating, with a magnified view showcasing the central portion revealing the individual pillars.}\label{structureG}
\end{figure}

As previously mentioned, to obtain SP-BICs, the structure must possess a high degree of symmetry. The simplest is axial symmetry, which in the case of a 1D periodic lattice translates into plane (mirror) symmetry. Consequently, a rectangular cross-section pillar grating deposited onto a planar substrate precisely fulfills the requirement, with axial symmetry perpendicular to the substrate interface.\\
\\
The substrate chosen for this application is lithium niobate (LN), selected for its set of non-linear properties that make it electromagnetically very important, mainly through modulation of its optical refractive index (Pockel's effect or pyroelectric effect) but also for SHG enhancement, in response to various external stimuli. The nano-structure consists of an array of pillars made of amorphous silicon (Si), a dielectric material with a high refractive index in the spectral range under consideration (NIR).\\
\\
To operate in the vicinity of $\lambda=1550~$nm, we performed numerical simulations using a customized FDTD code. Through the optimization of geometrical parameters, we aimed to achieve SP-BIC across the wavelength range of $\lambda\in[1200 - 1700]~$nm.
\\The geometrical characteristics of the structure after optimization, as depicted in Fig.~\ref{structureG}(a), are set as follows: the gap between pillars is $w=60~$nm, the height of pillars is $h=350~$nm, and the lattice period is $p=640~$nm. These specific parameters enable SP-BICs to be excited under different illumination conditions characterized by different values of $\theta, \psi, \phi$ (see Fig.~\ref{structureG}(a)), angles specifying the illumination properties (direction and polarization). Note that the ($x,y,z$) reference frame is aligned with the crystalline reference frame of the LN material, for consistency of notation. 

As shown in Fig.~\ref{structureG}(a), the LN substrate consists of an X-cut wafer, wherein the optical axis $z$ is oriented parallel to the Si-pillars. This configuration enables us to capitalize on the maximal electro-optical coefficient $r_{33}$ of the LN \cite{roussey2006electro,Lipson2018nanophotonic,RachelGrange} when the incident plane wave is polarized along the $z$ direction. Such alignment is imperative for achieving effective light-matter interaction. 

\section{Fabrication}
\begin{figure}[H]\centering
\includegraphics[clip,width=0.7\textwidth]{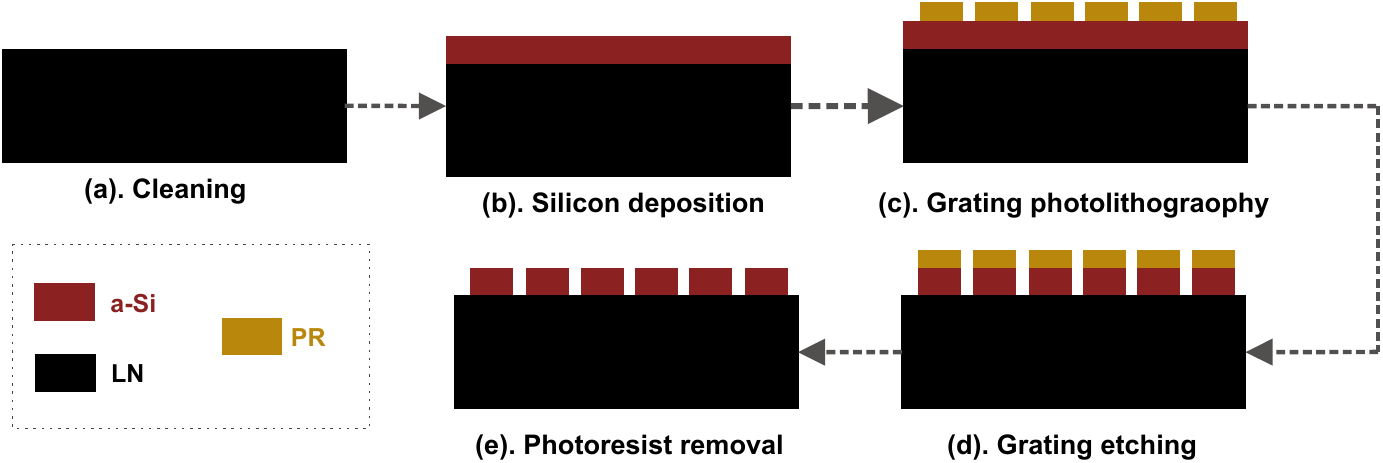}
\caption{Diagram and photograph of the optical bench used for transmission measurements. All elements are highlighted and indicated by yellow arrows.}\label{fabflow}
\end{figure}

Figure~\ref{structureG}(b) shows two scanning electron microscopy (SEM) images of the total manufactured structure (top), and a zoom-in (down) made over the central part revealing the high quality of the manufacturing process. The latter is outlined in the flowchart presented in Fig.~\ref{fabflow}. Initially, the X-cut LN wafer is cleaned using a piranha solution \cite{PiranhaSolution}. Subsequently, the Si layer is deposited at room temperature via the RF magnetron sputtering technique \cite{RF_sputtering}. Following this, a resist mask pattern of the Si nano-grating is formed atop the Si layer using e-beam lithography \cite{e_beam}, employing the CSAR-13 e-beam resist from Allresist. Next, the structure undergoes deep reactive ion etching (DRIE) \cite{DRIEfu2009} to generate the grating patterns, followed by the removal of the e-beam mask resist using NMP solution (N-methyl-2-pyrrolidone) at $70~^\circ$C for one hour.
The overall dimensions play a crucial role in exciting the SP-BIC \cite{djemaiai2024infinite}. Consequently, we fabricated a large-scale grating measuring $0.45\times0.45~\text{mm}^2$.

\section{Optical characterizations and numerical simulations}
\begin{figure}[H]\centering
\includegraphics[clip,width=0.9\textwidth]{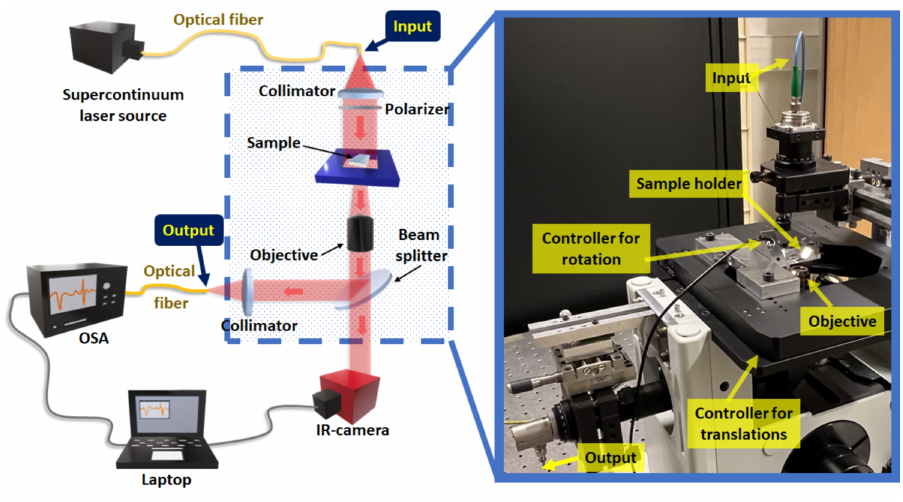}
\caption{Diagram and photograph of the optical bench used for transmission measurements. All elements are highlighted and indicated by yellow arrows.}\label{set-up}
\end{figure}

In order to measure the transmission properties of the structure, an optical bench was set up as shown in Fig.~\ref{set-up}, where all the essential optical elements are highlighted. Two piezoelectric motor controllers were used to facilitate simultaneous manipulation of the sample for translation (C-867KO58) and rotation (KIM001). The latter is needed to vary the angle of incidence ($\theta$), thereby facilitating the extrinsic symmetry breaking necessary for the excitation of the quasi-BICs, as we will see later.
The structure is exposed to a collimated beam emitted from a fibred continuum laser source (LEUKOS-SM20-OEM). These elements are integrated into an inverted optical microscope (Olympus IX51), which has been adapted to direct a portion of the transmitted signal to an Optical Spectrum Analyzer (OSA: MS9710B Anritsu) via an optical fiber. The remaining portion of the transmitted signal is used to control the position of the incident beam relative to the grating using a CDD IR camera (Goldeye CL-033 TEC$_\text{1}$).
\\The normalized transmission spectrum is obtained by dividing the signal transmitted through the structure by the signal transmitted only through the substrate; two measurements are therefore required per transmission spectrum.
As previously stated, simulations were conducted utilizing a custom FDTD code \cite{FDTD_computing}, enabling computation of the transmission spectrum under identical conditions as in experiment. The spatial discretization employed in the FDTD code comprises a cubic mesh with dimensions of $5\times5\times5~$nm$^3$ per cell. \\ It is important to note that despite the structure's invariance along one direction (in this case, the $Oz$ direction), a 2D simulation becomes inadequate due to the anisotropy of the LN material and/or in the case of azimuthal incidence ($\psi\neq0^\circ$). Consequently, a 3D-FDTD code with periodic boundary conditions is employed, considering only one period for simulations while accounting for LN anisotropy and oblique incidence.
\\To address this, we have developed a custom code by integrating anisotropy into the Split-Field Method \cite{belkhir2008three}, enabling consideration of oblique incidence in the case of an infinitely periodic anisotropic structures.

\section{Results}
Four distinct configurations were examined based on the direction and polarization of the incident plane wave. To distinguish them, we define three angles as depicted in Fig.~\ref{structureG}(a) : the azimuthal angle $\psi$, which specifies the angular orientation of the plane of incidence, $\phi$ related to the polarization angle ($\phi=0^\circ$ for TE and $\phi=90\circ$ for TM) in addition to the angle of incidence $\theta$.  
Hence, the four configurations studied, both numerically and experimentally, correspond to the following parameters:\\
\textbf{Configuration (a):} $\boldsymbol{(\psi,\phi)=(0^\circ,0^\circ)}$\\ 
\textbf{Configuration (b):} $\boldsymbol{(\psi,\phi)=(0^\circ,90^\circ)}$\\ 
\textbf{Configuration (c):} $\boldsymbol{(\psi,\phi)=(90^\circ,0^\circ)}$\\ 
\textbf{Configuration (d):} $\boldsymbol{(\psi,\phi)=(90^\circ,90^\circ)}$\\ 
However, for all configurations, the illumination is conducted from above the grating (air) with the angle of incidence $\theta$ ranging from $0$ to $30^\circ$. 
Notably, at normal incidence, configurations (a) and (b) are respectively equivalent to configurations (d) and (c). In the former case, the electric field aligns parallel to the Si-pillars ($z$ - axis), while in the latter, it becomes perpendicular. 
The optical refractive indices of LN are $n_x=n_y=n_o=2.211$ and $n_z=n_e=2.139$ \cite{refractiveindexpolyanskiy2024}. 
Yet, only configuration (d) represents an anisotropic case, while both configurations (c) and (d) require 3D numerical simulations.
The spectral positions of the Rayleigh anomalies are also calculated in order to better understand some features of the transmission spectra. These anomalies may interfere with other resonant modes, thereby altering their properties.
In the context of 1D periodic gratings, the spectral position ($\lambda_\text{A}$) of these anomalies, which are independent of polarization, are solutions of the equation defined below \cite{rayleigh1907dynamical}:
\begin{equation}
m^2\lambda_A^2+2pn_im\lambda_A\sin{\theta}\cos{\psi}+(n_i^2\sin^2{\theta}-\xi)p^2=0 
\label{anomalies}
\end{equation}
Here, $m$ denotes the order of diffraction in the periodicity direction ($y$-axis), $n_i$ is the optical refractive index of the medium of incidence (air), $n_t$ is the optical refractive index of the medium of transmission (LN), $\xi$ is equal to $n_i^2$ or $n_t^2$ for Rayleigh anomalies occurring in the incident or transmission medium respectively, and $p$ is the period of the grating. 
In the chosen range of study, only the Rayleigh anomalies occurring in the transmission medium are visible. The solutions to Eq.~\ref{anomalies} can therefore be expressed as follows: 
\\When $\psi=0^\circ$, there exist two solutions:
$$
\lambda_A = \left\{
    \begin{array}{ll}
       -\frac{p}{m}\times~(n_i\sin{\theta}+n_t) \ & \text{for~}m<0 \\
       -\frac{p}{m}\times~(n_i\sin{\theta}-n_t) \ & \text{for~}m>0
    \end{array}
\right.
$$
\\And when $\psi=90^\circ$, there exists only one solution:
$$
\lambda_A=\frac{p}{m}\times~\sqrt{n_t^2-n_i^2\sin^2{\theta}}
$$
All the materials used behave like pure dielectrics in the near-infrared spectrum, with no absorption, enabling the quantification of the optical response of the structure, whether in transmission or reflection. Consequently, only the transmission coefficient ($T$) will be presented.
\\In addition, we introduce the following parameters for the quantification of the obtained results. Firstly, the quality-factor, denoted as $Q$, is determined by the ratio of the resonance wavelength $\lambda_0$ to the full width at half-maximum (FWHM) $\Delta\lambda$ of a resonant peak. For resonances exhibiting Fano asymmetric line shapes, we consider the FWHM as the absolute difference between the spectral positions of the pic and the dip of the resonance. Secondly, we define the extinction ratio (ER) of the normalized transmission spectra, expressed by $\text{ER}=\frac{T_{max}-T_{min}}{T_{max}}$ where $T_{max}$ and $T_{min}$ represent the maximum and minimum normalized transmission intensities, respectively. This parameter serves to evaluate the efficiency of light modulation as will be seen later. 
Lastly, we introduce the normalized electric and magnetic field intensities ($I_e$ and $I_m$), calculated as the ratio of the intensity in the presence of both the grating and the substrate to the intensity in the presence of only the substrate. These quantities provide insight into the interaction dynamics between incident light and the grating-substrate system.
\subsection{\underline{Configuration (a)}: $\boldsymbol{(\psi,\phi)=(0^\circ,0^\circ)}$}
\begin{figure}[H]\centering
\lineskip=0pt
{\includegraphics[clip,width=1\textwidth]{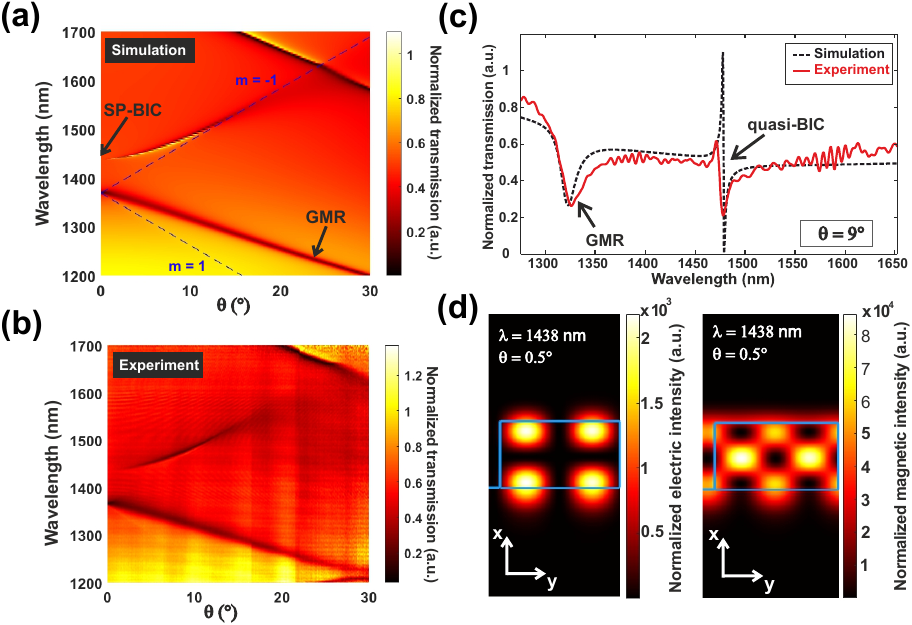}}
\caption{(a) Transmission angular diagram obtained by FDTD numerical simulations. The dashed blue lines represent the positions of the Rayleigh anomalies for both the $m=1$ and $m=-1$ diffraction orders. (b) Transmission angular diagram obtained from experimental measurements. (c) The transmission spectra acquired from simulation (black dashed line) and experiment (red solid line) at the angle of incidence $\theta=9^\circ$. (d) Distributions of the normalized electric and magnetic intensities obtained from FDTD simulations at the spectral position of the quasi-BIC $\lambda=1438~$nm for an incidence angle of $\theta=0.5^\circ$.}
\label{figure-a}
\end{figure}

In this configuration, the plane of incidence ($xy$ plane) is perpendicular to the direction of the Si-pillars ($z$-axis). Moreover, the E-field is oriented along the $z$-axis ($\phi=0^\circ$), parallel to the Si-pillars. This corresponds to a conventional TE polarization case with no depolarization, as the plane of incidence is perpendicular to the direction of invariance (the $z$-axis). Consequently, the numerical simulations are entirely two-dimensional, reducing the electromagnetic field to only the ($H_{\text{x}}$, $H_{\text{y}}$, $E_{\text{z}}$) components. \\
Figure~\ref{figure-a} illustrates both numerical (a) and experimental (b) results of the transmission angular diagrams. Note that the experimental ones are unprocessed raw data. Additionally, the Rayleigh anomalies of the diffracted first orders ($m=\pm1$) are depicted by dashed blue lines in Fig.~\ref{figure-a}(a). There is excellent agreement between experience and theory (Fig.~\ref{figure-a}(a) and \ref{figure-a}(b)) in terms of excited modes and resonance spectral positions. 
\\Small differences can have several origins, such as the finite size of the structure and/or the illumination beam, and also the unavoidable manufacturing defects. 
Under normal incidence ($\theta=0^\circ$), only the Guided Mode Resonance (GMR) is observable at the wavelength of $\lambda=1368~$nm. However, a slight variation of the angle of incidence $\theta$ leads to a break, even slight, in the symmetry of the structure relative to the $x$-axis. Consequently, we observe the emergence of a Fano-shaped quasi-BIC above the $m=-1$ Rayleigh anomaly. This reveals the presence of a Symmetry Protected - Bound state in the Continuum (SP-BIC) at the point of high symmetry ($\theta=0^\circ$).
\\As the angle $\theta$ increases, the GMR experiences a blue shift, whereas the quasi-BIC undergoes a red shift. For example, Fig.~\ref{figure-a}(c) illustrates the normalized transmission spectra obtained from both numerical simulation (blue dashed line) and experimental measurement (red solid line) at the angle of incidence $\theta=9^\circ$. Within the same figure, one can observe the presence of the two resonances:  the GMR one located around $\lambda=1330~$nm and the quasi-BIC one at $\lambda=1477~$nm. Beyond $\theta\approx13^\circ$, the SP-BIC couples destructively with the Rayleigh anomaly of order $m=-1$. This leads to an increase in its quality factor before it disappears completely, giving rise to an accidental BIC \cite{accidentalBIC2013observation}.

\begin{figure}[H]\centering
\lineskip=0pt
{\includegraphics[clip,width=0.8\textwidth]{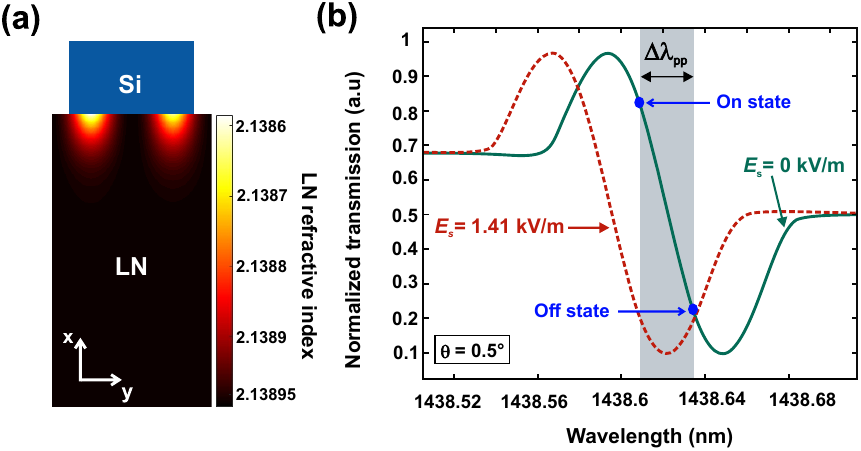}}
\caption{(a) Spatial distribution of the LN refractive index induced by the application of an external electric field $E_s=1.41~$kV/m. (b) The normalized transmission response of the structure under an external electric field $E_s$ (cases of $E_s$=1.41kV/m (red dashed line) and $E_s=0$~kV/m (green solid line)).}
\label{appte0}
\end{figure}

To go further in the quasi-BIC properties, we examine a small angle of incidence, for instance, $\theta=0.5^\circ$ for which the quasi-BIC resonates at a wavelength of $\lambda_{num}=1438~$nm, exhibiting a substantial Q-factor of $Q_{num}=10^6$. We numerically demonstrate that this resonance is accompanied by a significant enhancement of both electric and magnetic field intensities, with values of $I_e=2.10^3$  and $I_m=8.10^4$, respectively, as illustrated in Fig.~\ref{figure-a}(d).
Based on these results, the quasi-BIC could be ideally suited for a wide range of integrated photonics applications, particularly in areas such as optoelectronics \cite{Zhang:oe22,hoblos2020excitation} and nonlinear optics \cite{baida2023giant,li:lsa22}.
The fact that the electric field of the electromagnetic wave is polarised along the LN's crystalline axis ($z$-axis) makes the configuration highly compatible with electro-optical or pyroelectric applications. Indeed, the largest electro-optic (EO) coefficient of the LN is $r_{33}=31~$pm/V \cite{LNpropertieswong2002}, which is mainly involved in the Pockel's effect in this case. The obtained variation in the refractive index is given by Eq. 2 \cite{hoblos2020excitation} below.
\begin{eqnarray}
    \Delta n(x,y)&=&-\frac{1}{2}n_e^3r_{33}f(x,y)^2E_s\\
    E_s&=&-\frac{1}{\varepsilon_0\varepsilon_r}p_t\Delta\text{T}
     \label{equation eo}
\end{eqnarray}
Here, $E_s$ denotes the external applied electric field magnitude, $\varepsilon_0$ and $\varepsilon_r$ are, respectively, the vacuum and the LN relative dielectric constants equal to $\varepsilon_0=8.854\times10^{-12}~$F/m, and $\varepsilon_r=28.7$. The function $f(x,y)$ represents the local optical field factor computed at each FDTD cell. It quantifies the electric field amplitude in the presence of the grating relative to the field amplitude in the absence of the grating. 
However, this refractive index modification can also be induced by a change in temperature (pyroelectric effect) which, through induced polarization, results in the generation of an electric field within the LN crystal. A relationship linking the two effects is given by Eq. 3 where $p_t$ is the pyroelectric coefficient of the LN ($p_t=-6\times10^{-5}~$Cm$^{-2}$K$^{-1}$), and $\Delta$T corresponds to the temperature variation in kelvin.
Let consider an external electric field applied to the grating by integrating two electrodes deposited on the LN substrate at both ends of the grating along the $z$-axis. We consider a distance of $500~\mu$m between the electrodes to cover the area of the grating since, as mentioned above, the total dimension of the grating along the $z$- and $y$-axes is $450\times450~\mu$m$^2$. Figure~\ref{appte0}(a) depicts the local variation of the refractive index of LN induced by an external electric field of $E_s=1.41~$kV/m corresponding to voltage of $0.7$~V applied on the electrodes, or by temperature variation of $\Delta T=6~mK$. 
 
The modification of the refractive index is then calculated through Eq. 2 for each FDTD-cell of the mesh and incorporated into the FDTD simulations to estimate the effect on the transmission response. Figure \ref{appte0}(b) presents the two spectra corresponding to the passive structure when $E_s=0~V/m$ (green solid line) and to an applied electric field of $E_s=1.14~$kV/m (dashed red line). As expected, due to the large local enhancement of the electric field in the LN just above the Si-pillars, the refractive index of the latter is enough decreased allowing a large blue-shift ($\Delta\lambda_{res}=5.26\times10^{-2}~$nm) of the SP-BIC resonance. 

When the aim is to linearly modulate the intensity of the transmitted signal, the latter must change uniformly from an off-state to an on-state. These states, depicted by blue circles in Fig.~\ref{appte0}(b), denote the lowest and the highest values of the normalized transmission at the quasi-BIC resonance's linear segment. This is achieved through a wavelength shift of $\Delta\lambda_{\text{pp}}=2.24\times10^{-3}~$nm corresponding to a an extinction ratio of $ER=77$\% and a peak-to-peak drive voltage $V_\text{pp}=0.7$~V as shown on Fig.~\ref{appte0}(b). 
\\Furthermore, we can assess the thermal sensitivity of the structure by exploiting the LN intrinsic pyroelectric property given by the Eq.~3. as $S_t=\Delta\lambda/\Delta T = 0.37\times10^9$~nm/K.
This large thermal sensitivity can be exploited for highly accurate temperature detection in the milli-Kelvin range, better than the state of the art \cite{highEO2024,temperatureqiu2016fano,temperaturesensorpark2014double,temperaturesensitivitypeng2012}, but can also play a negative role in other applications, then requiring correction of thermal drifts.

\subsection{\underline{Configuration (b)}: $\boldsymbol{(\psi,\phi)=(0^\circ,90^\circ)}$}
\begin{figure}[H]\centering
{\includegraphics[clip,width=0.97\textwidth]{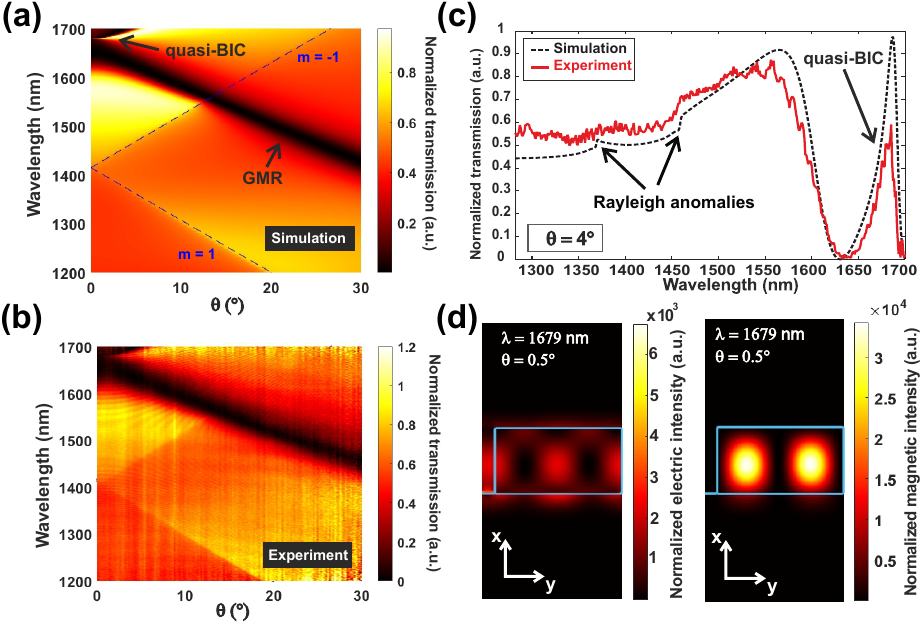}}
\caption{(a) Transmission angular diagram derived from FDTD numerical simulations. The dashed blue lines represent the spectral positions of the Rayleigh anomalies for both the $m=1$ and $m=-1$ diffraction orders. (b) Transmission angular diagram obtained from experimental measurements. (c) The transmission spectra acquired from simulation (black dashed line) and experiment (red solid line) at the angle of incidence $\theta=4^\circ$. (d) Distributions of the normalized electric and magnetic intensities obtained from FDTD simulations at the spectral position of the quasi-BIC $\lambda=1679~$nm for an incidence angle of $\theta=0.5^\circ$.}\label{figure-b}
\end{figure}

\begin{figure}[H]\centering
\lineskip=0pt
\includegraphics[clip,width=0.9\textwidth]{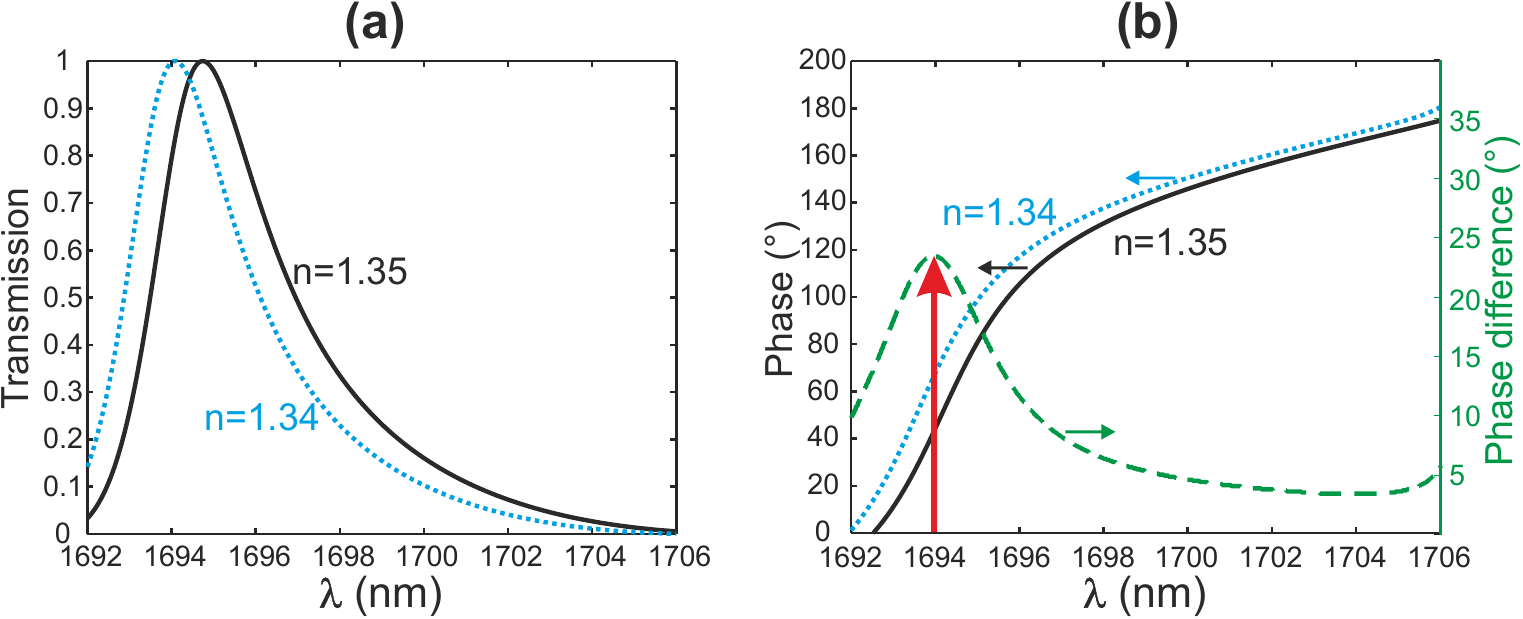}
\caption{(a) Transmission spectra in the region of the quasi-BIC of the structure with two different values of the refractive index $n_l$ of the liquid superstate ($n_l=1.34$ in dotted blue line and $n_l=1.35$ in solid black line).  (b) Corresponding phase changes of the transmitted electric fields for the two values ($n_l=1.34$ in dotted blue line and $n_l=1.35$ in solid black line). The phase difference is plotted in dashed green line with a maximum value indicted by the vertical red arrow corresponding to 23.49$^\circ$ at $\lambda=1693.95$~nm.}
\label{spectrenl}
\end{figure}

In this configuration, the plane of incidence remains the same as in configuration (a), namely perpendicular to the Si-pillars. However, the polarization state changes to a transverse magnetic (TM) with $\phi=90^\circ$. Consequently, the electromagnetic field components are now characterized by ($E_x, E_y, H_z$), where the electric field aligns with both ordinary axes ($x$ and $y$), resulting in an isotropic study case. Numerical simulations and experimental measurements of the transmission angular diagram are presented in Fig.~\ref{figure-b}(a) and Fig.~\ref{figure-b}(b) respectively. As before, the Rayleigh anomalies are also indicated by the blue dashed lines.
In both diagrams, we observe the emergence of a quasi-BIC above the Rayleigh anomalies when the angle of incidence increases leading to its apparition at approximately $\lambda=1678~$nm. The numerical and experimental normalized transmission spectra at an angle of $\theta=4^\circ$ are illustrated in Fig.~\ref{figure-b}(c), where the quasi-BIC is located at a wavelength of $\lambda=1686$~nm. In contrast to configuration (a), the quality factor of the SP-BIC decreases considerably with small symmetry breaks giving rise to a broadened resonance. In order to exploit it in detection, modulation, or non-linear signal generation applications, it would be necessary to operate at small incidence angles.
Therefore, to quantify the confinement of the electromagnetic field, we consider an angle of incidence $\theta=0.5^\circ$ for which the mode appears at $\lambda=1679~$nm with a Q-factor of $Q=892$. As depicted in Fig.~\ref{figure-b}(d), the electric intensity associated with the mode is predominantly enhanced at the surface of the LN substrate within the air cavities, reaching a maximum of $I_e=6\times10^3$, along with an enhancement in magnetic intensity, mainly confined inside the Si-pillars, peaking at $I_m=3\times10^4$. In particular, we observe a strong increase in magnetic intensity, probably due to the excitation of a magnetic dipolar resonance of the Si-pillars, which considerably increases the MLDOS (local density of magnetic states), opening the way to magneto-optical applications with non-absorbing structures \cite{mivelle:nl18,KOSHELEV2019836} other than metallic nano-antennas \cite{Darvishzadeh:ol19}. 
As mentioned above, the electric field at resonance is significantly enhanced at the substrate interface between two Si-pillars in the gap region. This could be exploited to build a bio-sensor or refractive index detector. In this case, the superstrate is no more air but a liquid (water or blood) with refractive index $n_l$. FDTD simulations are then done by considering two fairly close values of $n_l$ in order to evaluate the refractive index sensitivity of the structure in the vicinity of the quasi-BIC. Figure \ref{spectrenl}(a) shows the obtained transmission spectra in the spectral region of the quasi-BIC resonance for $n_l=1.34$ and $n_l=1.35$ when the angle of incidence from air on the liquid interface is fixed to $\theta=0.75^\circ$. We can clearly see that the transmission peak is slightly shifted by $\Delta\lambda=0.7$~nm corresponding to a weak sensitivity of 70nm/RIU, not sufficiently relevant for a refractive index detection even if the Q-factor of the resonance is almost high ($Q=490$) . Nevertheless, phase detection instead of intensity detection can, in certain configurations, greatly improve sensitivity \cite{phasekabashin2009}. Figure \ref{spectrenl}(b) presents the corresponding phase change induced during the transmission of light through the structure for the same two values of the refractive index (in dotted blue line for $n_l=1.34$ and in solid black line for $n_l=1.35$). At $\lambda=1693.95$~nm, the phase difference, plotted in green dashed line, is maximum and is worth $23.48^\circ$ which gives a good phase sensitivity of $2348^\circ/$RIU. This value is considerably high \cite{indexsensing2019symmetric,sabrina2021high}, making the structure an excellent candidate for a variety of applications like label-free bio-sensing, bio-imaging, and optical filters.

\subsection{\underline{Configuration (c)}: $\boldsymbol{(\psi,\phi)=(90^\circ,0^\circ)}$}
\begin{figure}[H]\centering
\lineskip=0pt
{\includegraphics[clip,width=0.9\textwidth]{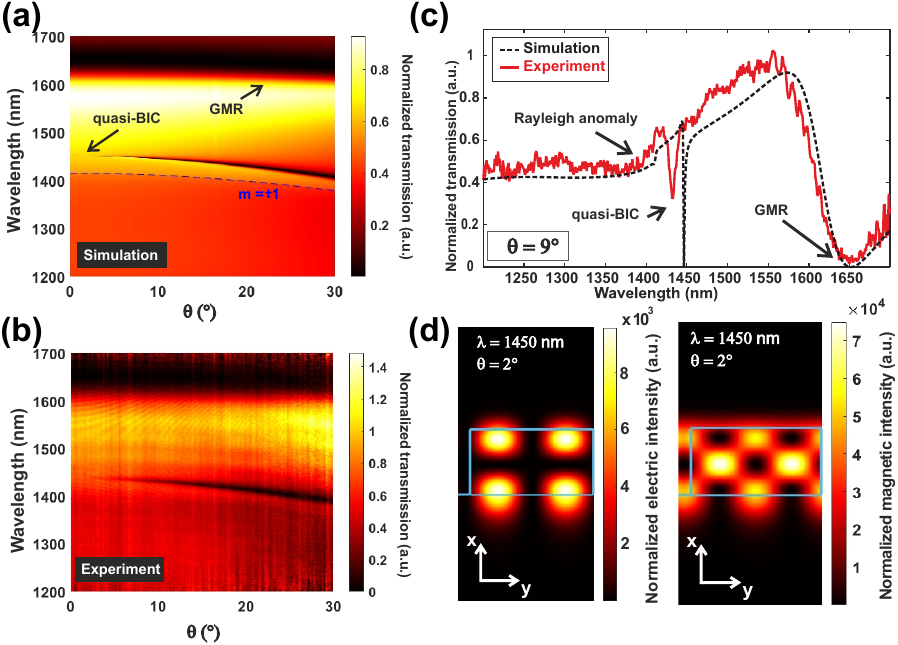}}
\caption{(a) Transmission angular diagram derived from FDTD numerical simulations. The dashed blue line represents the spectral positions of the Rayleigh anomaly for the first diffracted order ($m=\pm1$). (b) Transmission angular diagram obtained from experimental measurements. (c) The transmission spectra obtained from simulation (black dashed line) and experiment (red solid line) at the angle of incidence $\theta=9^\circ$. (d) Distributions of the normalized electric and magnetic intensities obtained from FDTD simulations at the spectral position of the quasi-BIC $\lambda=1450$~nm for an incidence angle of $\theta=2^\circ$.}
\label{figure-c}
\end{figure}

In this configuration, the azimuthal angle $\psi$ is set to $90^\circ$, corresponding to a plane of incidence parallel to the $xz$ plane and aligned with the Si-pillars. The polarization is set to \emph{TE} ($\phi=0^\circ$) so that the electric field remains perpendicular to the Si-pillars, whatever the angle of incidence. By diffraction, and due to invariance along the z direction, this electric field will give rise to a field only located in the $xy$ plane, making the theoretical problem completely isotropic. In addition, in this configuration, as in the (d) one, changing the angle of incidence induces a very slight change in illumination conditions, so this case is rarely considered in the literature.
\\The numerical and experimental transmission angular diagrams are illustrated, respectively, in Fig.~\ref{figure-c}(a) and \ref{figure-c}(b). Once again, the simulations lead to results very faithful to those obtained experimentally (see Fig.~\ref{figure-c}(c)), showing the excitation of one SP-BIC mode and one GMR. These two modes are the same as the ones obtained in configuration (a) for the quasi-BIC, and in configuration (b) for the GMR. In fact, at normal incidence, configuration (c) is completely identical to configuration (b) while the occurrence of quasi-BIC is due to the diffraction-induced depolarization at $\theta\ne0^\circ$ and more precisely the emergence of a non zero x-component of the electric field. As expected, the weak extrinsic symmetry break ($\theta\neq0$) reveals this quasi-BIC at the same wavelength as in configuration (a) above the Rayleigh anomaly, as depicted in Fig.~\ref{figure-c}(c) for $\theta=9^\circ$. The quasi-BIC resonance exhibits a good quality factor over a wide range of angle of incidence (up to $Q_{num}=0.94\times10^6$ for $\theta=0.5^\circ$), making this configuration more robust with regard to this experimental condition of illumination.

At the aforementioned quasi-BIC resonance, the electromagnetic field distributions (Fig.~\ref{figure-c}(d) ) are obviously similar to those of configuration (a) (Fig.~\ref{figure-a}(d)), but with a slightly higher enhancement as shown on Fig.~\ref{figure-c}(d). The normalized electric and magnetic field intensities reach $I_e=9\times10^3$ and $I_m=8\times10^4$ respectively for $\theta=2^\circ$ while similar enhancement factors need a smaller angle of incidence ($\theta=0.5^\circ$) in the configuration (a). 
Nevertheless, even if the origin of the quasi-BIC is the same (Mie-type magnetic resonance) for both configurations (a) and (c), the spectral behavior with respect to the variation of the angle of incidence is different. In fact, for the (c) configuration, increasing the grating's asymmetry induces a blue-shift to the quasi-BIC while it red-shifts in configuration (a).
Furthermore, the potential applications of the quasi-BIC generated in this configuration are very similar to those proposed in configuration (a) due to their similar confinement properties. However, at high incidence angles ($\theta>4^\circ$), the quasi-BICs of the two configurations occupy different spectral positions and can therefore be used simultaneously for the same or different applications, depending on the desired performance of each.

\subsection{\underline{Configuration (d)}: $\boldsymbol{(\psi,\phi)=(90^\circ,90^\circ)}$}
\begin{figure}[H]\centering
\lineskip=0pt
{\includegraphics[clip,width=1\textwidth]{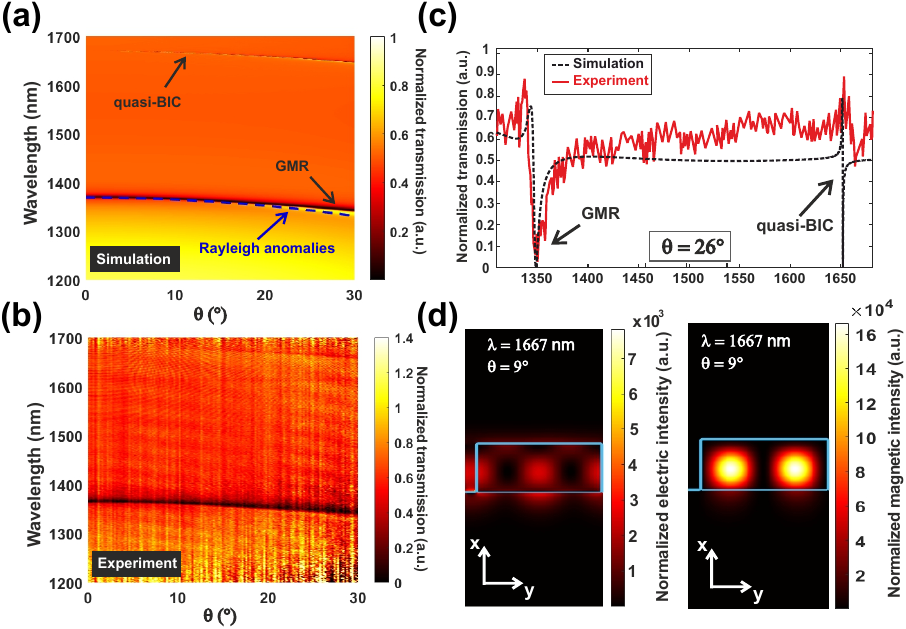}}
\caption{(a) Transmission angular diagram derived from FDTD numerical simulations. The dashed blue line represents the spectral positions of the Rayleigh anomaly for the first diffracted order ($m=\pm1$). (b) Transmission angular diagram obtained from experimental measurements. (c) The transmission spectra acquired from simulation (black dashed line) and experiment (red solid line) at the angle of incidence $\theta=26^\circ$. (d) Distributions of the normalized electric and magnetic intensities obtained from FDTD simulations at the spectral position of the quasi-BIC $\lambda=1667$~nm for an incidence angle of $\theta=9^\circ$.}\label{figure-d}
\end{figure}

In this configuration, both angles $\psi$ and $\phi$ (defined in Fig.~\ref{structureG}(a)) are set to $90^\circ$. Therefore, the plane of incidence is the $xz$ plane as in configuration (c) but the polarization state is \emph{TM}. Consequently, the electric field is parallel to the same plane $xz$ and has now two components along both the ordinary and extraordinary axes ($x$ and $z$). The 3D-FDTD simulations must take into account this anisotropy of the configuration. 
The numerical and experimental transmission angular diagrams are presented in Fig.~\ref{figure-d}(a) and \ref{figure-d}(b). As before, there is a good agreement between the two results.
In Fig.~\ref{figure-d}(a), the dashed blue line identifies the Rayleigh anomaly corresponding to the first diffracted order ($m=\pm1$) calculated using Eq. 1. Both diagrams emphasise the presence of two modes. However, under normal incidence, this configuration is identical to configuration (a), where a single mode exists above the Rayleigh anomalies, which is actually the GMR at $\lambda=1370$~nm. The second mode, rather more difficult to see experimentally, only appears at $\theta\ne0^\circ$, presenting an exceptional quality factor of $Q_{num}=1.3\times10^6$ at $\theta=0.5^\circ$, indicating a quasi-BIC state. This reveals the presence of an SP-BIC at the highest symmetry point (here $\Gamma$). 
As with configuration (c), quasi-BIC maintains a significant quality factor even at higher angles of incidence. For example, at $\theta=26^\circ$ its value remains significant, reaching $Q_{num}=0.57\times10^4$, with a perfectly asymmetric Fano-shape, making this configuration of great potential for lasers and Q-switch applications.
Moreover, this quasi-BIC is the same as the one excited in configuration (b) at around $\lambda=1679$~nm and is due to depolarization, which reveals a y-component that remains fairly weak whatever the angle of incidence. Figure~\ref{figure-d}(c) illustrates the electric and magnetic intensity distributions for $\theta=9^\circ$. As expected, these distributions are perfectly identical to those of the quasi-BIC achieved in configuration (b). However, the quasi-BIC generated in this configuration offers better confinement thanks to its high Q-factor, and has the advantage of being more robust with respect to the angle of incidence.

By observing the symmetry of the electric field distribution in Fig.~\ref{figure-d}(d), we can see that it is intimately dependent on the shape of the Si-pillar. A mechanical vibration of this pillar is therefore likely to disrupt the excitation of this quasi-BIC by shifting its excitation wavelength. This will modulate the intensity of the transmitted signal. Consequently, we investigate the structure sensitivity to such mechanical vibration, particularly relevant for acousto-optical applications \cite{yu2020acousto}. 

\begin{figure}[H]\centering
\lineskip=0pt
{\includegraphics[clip,width=0.9\textwidth]{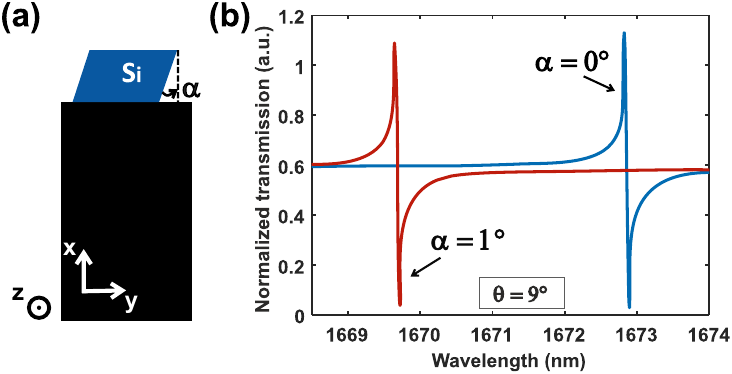}}
\caption{(a) Diagram of a structure subjected to bending of the pillars at an angle $\alpha$. (b) Comparison of the spectral position of the quasi-BIC when $\alpha=1^\circ$ (red dashed line) versus $\alpha=0^\circ$ (green solid line) at the angle of incidence $\theta=9^\circ$. }
\label{apptm90}
\end{figure}

A fairly simple approach is to consider a uniform tilt of the Si-pillar as shown in Fig.~\ref{apptm90}(a) where $\alpha$ denotes the tilt angle of the pillar with respect to the vertical direction. To estimate the influence of pillar bending, we calculated the transmission spectrum for $\alpha=1^\circ$ (corresponding to a maximum displacement of $6.1$~nm for the top of the pillar) and compared it to that of the initial structure ($\alpha=0^\circ$). Both spectra are shown in Fig.~\ref{apptm90}(b) where the quasi-BIC undergoes a shift of $\Delta\lambda=3.3$~nm leading to a sensitivity of $\cfrac{\Delta\lambda}{\Delta \alpha}=3.3$~nm/deg. 
We believe that this bending is made possible by a surface acoustic wave (SAW), directed along the $y$-axis, and induced by an interdigital comb arranged alongside the structure, on the surface of the LN piezoelectric substrate.

\section{Conclusion}
\begin{figure}[H]\centering
\lineskip=0pt
{\includegraphics[clip,width=0.7\textwidth]{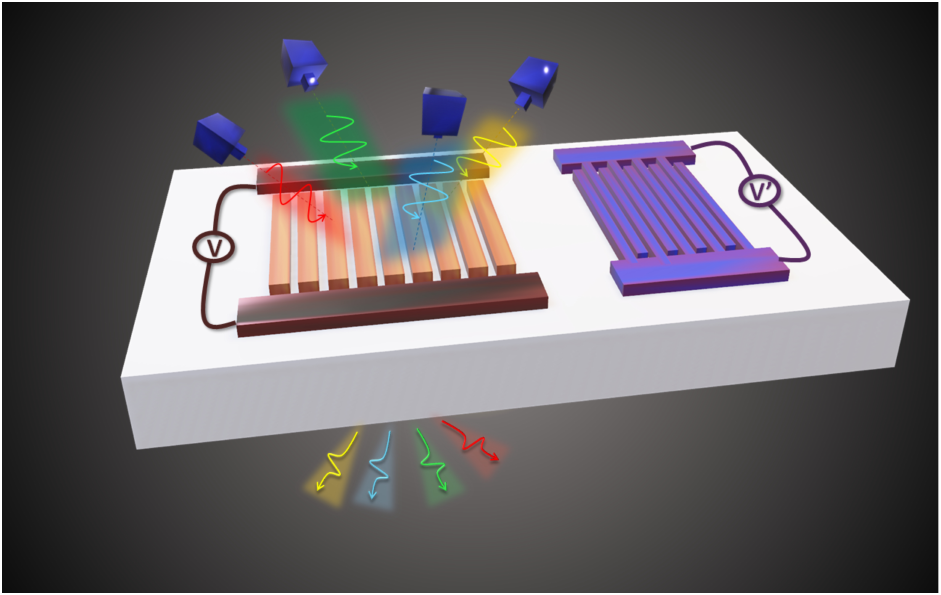}}
\caption{
Artistic illustration showing the versatility of the proposed configuration.}\label{multitaskingG}
\end{figure}

We have demonstrated the ability to excite quasi-BIC modes by symmetry breaking, using four distinct configurations within the same simple-to-fabricate structure which doesn't involve lithium niobate structuring. By breaking the lattice symmetry, leakage to the SP-BIC occurs, resulting in a resonant mode (quasi-BIC) within the lattice with a particularly high Q factor. Numerical simulations and experimental measurements show excellent agreement, confirming the reliability of the proposed structure. 
In each configuration, the quasi-BIC investigation was able to demonstrate the quality of the electromagnetic field confinement. We note that of the four configurations, only two SP-BICs exist, but that their coupling to the continuum can occur in two distinct ways. For example, while configurations (a) and (c) differ in various aspects such as plane of incidence, LN's refractive index and levels of symmetry breaking, they both produce quasi-BICs with similar characteristics; comparable wavelengths at low angles of incidence ($\theta\leq0.5^\circ$) and similar field distributions, enabling its exploitation under different experimental conditions as schematically shown in Fig. \ref{multitaskingG}. Similarly, configurations (b) and (d) show similarities in their quasi-BICs excited at low angles of incidence ($\theta\leq0.5^\circ$), indicating the common origin of the SP-BIC. However, the four generated quasi-BICs show distinct responses to increasing asymmetry via the angle of incidence. 
In addition to the compact size of the structure and the etch-free of the LN, all quasi-BICs offer significant light confinement, making the proposed structure an efficient miniaturized device for in chip-scale applications in the NIR.
\section*{Acknowledgments}
Computations have been partially performed on the supercomputer facilities of the "M\'esocentre de calcul de Franche-Comt\'e". This work has been achieved in the frame of the EIPHI Graduate school (contract "ANR-17-EURE-0002) and supported by French RENATECH network through its FEMTO-ST technological facility. The authors would like to thank Prof. Jacquot Maxime for providing the inverted microscope around which the optical bench was built.

\section*{Disclosures}
The authors declare no conflicts of interest.

\section*{Data Availability}
Data underlying the results presented in this paper are not publicly available at this time but may be obtained from the authors upon reasonable request.


-----------------------------

\bibliography{main}

\end{document}